\documentclass[11pt]{article}
\usepackage{graphicx}
\usepackage{amsmath}
\usepackage{amssymb}
\usepackage{amsxtra}

\title{Novel Perspectives from Light-Front QCD, Super-Conformal Algebra, \\ and Light-Front Holography}
\author{Stanley J. Brodsky\\SLAC National Accelerator Laboratory\\ Stanford University\\sjbth@slac.stanford.edu}

\begin{document}
\maketitle

\begin{abstract}

Light-Front Quantization --  Dirac's ``Front Form" -- provides a physical, frame-independent formalism for hadron dynamics and structure.  Observables such as structure functions, transverse momentum distributions, and distribution amplitudes are defined from the hadronic LFWFs.   One obtains new insights into the hadronic mass scale, the hadronic spectrum, and the functional form of the QCD running coupling in the nonperturbative domain using light-front holography.  In addition, superconformal algebra leads to remarkable supersymmetric relations between mesons and baryons.
I also discuss evidence that the antishadowing of nuclear structure functions is non-universal;  i.e.,  flavor dependent, and why shadowing and antishadowing phenomena may be incompatible with the momentum and other sum rules for the nuclear parton distribution functions.

\end{abstract}

\section{Light-Front Wavefunctions and QCD}

Measurements of hadron structure -- such as the structure functions determined by  deep inelastic lepton-proton scattering (DIS) -- are analogous to a flash photograph: one observes the hadron at fixed 
$\tau=  t+z/c$ along a light-front, not at a given instant of time $t$.  The underlying physics follows from the 
the light-front  wavefunctions (LFWFs)  
$\psi_n(x_i,  \vec k_{\perp i },  \lambda_i )$ with 
$x_i = {k^+_i\over P^+} = {k^0_i+k^z_i\over P^0+P^z}, \sum^n_i x_1 =1, \sum^n_i \vec k_{\perp _i} =\vec 0_\perp$  and spin projections $\lambda_i$.  The LFWFs are the Fock state projections of the eigenstates of the QCD LF Hamiltonian $H_{LF} |\Psi> = M^2|\Psi>$~\cite{Brodsky:1997de}, where the LF Hamiltonian is the light-front time evolution operator defined directly from the QCD Lagrangian. 
One can avoid ghosts and longitudinal  gluonic degrees of freedom by choosing.to work in the light-cone gauge  $A^+ =0$.  
The LFWFs are boost invariant; i.e., they are independent of the hadron's momentum $P^+ =P^0 +P^z, \vec P_\perp.$
This contrasts with the wavefunctions defined at a fixed time $t$ -- the Lorentz boost of an instant-form wavefunction is much more complicated than a 
Melosh transform~\cite{Brodsky:1968ea} -- even the number of Fock constituents changes under a boost.  
Current matrix element such as form factors are simple overlaps of  the initial-state and final-state LFWFs, as given by the 
Drell-Yan West formula~\cite{Drell:1969km, West:1970av,Brodsky:1980zm}. There is no analogous formula for the instant form, since one must take into account the coupling of the external current to connected vacuum-induced currents.
Observables such as structure functions, transverse momentum distributions, and distribution amplitudes are defined from the hadronic LFWFs. 
Since they are frame-independent, the structure functions measured in DIS are  the same whether they are measured in an electron-proton collider or in a fixed-target experiment where the proton is at rest.     There is no concept of length contraction of the hadron or nucleus at  a collider -- no collisions of  ``pancakes" --   since the observations  of the collisions of the composite hadrons are made at fixed $\tau$, not  at fixed time.    The dynamics of a hadron is not dependent on the observer's Lorentz frame.  

The LF Heisenberg equation can in principle be solved numerically by matrix diagonalization  using ``Discretized Light-Cone  Quantization" (DLCQ)~\cite{Pauli:1985pv} where anti-periodic boundary conditions in 
$x^-$ render the $k^+$ momenta  discrete  as well as  limiting the size of the Fock basis.  In fact, one can easily solve 1+1 quantum field theories such as QCD$(1+1)$ ~\cite{Hornbostel:1988fb} for any number if colors, flavors and quark masses.  Unlike lattice gauge theory, the nonpertubative DLCQ analysis is in Minkowski space, is frame-independent and is free of fermion-doubling problems.   A new method for solving nonperturbative QCD ``Basis Light-Front Quantization" (BLFQ)~\cite{Vary:2014tqa},  uses the eigensolutions of a color-confining approximation to QCD (such as LF holography ) as the basis functions,  rather than the plane-wave basis used in DLCQ.  The LFWFs can also be determined from covariant Bethe-Salpeter wavefunction by integrating over $k^-$~\cite{Brodsky:2015aia}.  

Factorization theorems and  DGLAP and ERBL evolution equations can be derived using the light-front Hamiltonian formalism~\cite{Lepage:1980fj}.  In the case of an electron-ion collider, one can represent the cross section for e-p colisions as a convolution of the hadron and virtual photon structure functions times the subprocess cross-section in analogy to hadron-hadron colisions.   This nonstandard description of $\gamma^* p \to X$ reactions  gives new insights into electroproduction physics -- physics not apparent using the usual usual infinite momentum frame description, such as the dynamics of heavy quark-pair production.  
intrinsic heavy quarks also play an important role~\cite{Brodsky:2015uwa}.

In the case of $e p \to e^\prime X$, one can consider the collisions of the confining  QCD flux tube appearing between the $q$ and $\bar q$  of the virtual photon with the flux tube between the quark and diquark of the proton.   Since the $q \bar q$ plane is aligned with the scattered electron's plane, the resulting ``ridge"  of hadronic multiplicity produced from the $\gamma^* p$ collision will also be aligned with the scattering plane of the scattered electron.  The virtual photon's flux tube will also depend on the photon virtuality $Q^2$, as well as the flavor of the produced pair arising from $\gamma^* \to q \bar q$.  The resulting dynamics~\cite{Brodsky:2014hia} is  a natural extension of the flux-tube collision description of the ridge produced in $p-p$ collisions~\cite{Bjorken:2013boa}.

\section{Color Confinement and Supersymmetry in Hadron Physics from LF Holography}

\begin{figure}
 \begin{center}
\includegraphics[height= 12cm,width=15cm]{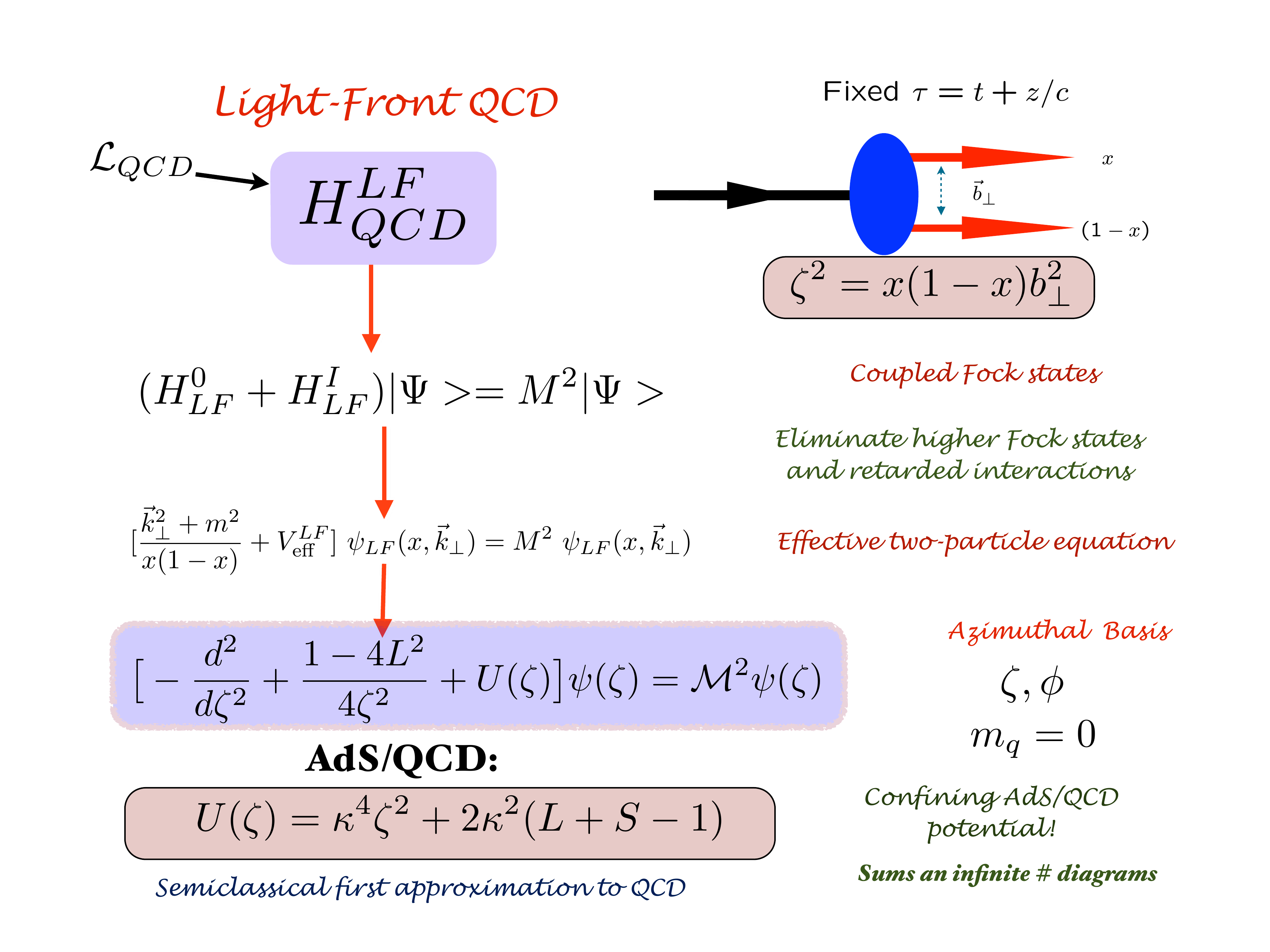}
\end{center}
\caption{Derivation of the Effective Light-Front Schr\"odinger Equation from QCD.  As in QED, one reduces the LF Heisenberg equation $H_{LF}|\Psi >   = M^2 |\Psi>$ 
to an effective two-body eigenvalue equation for $q \bar q$ mesons by systematically eliminating higher Fock states. One utilizes the LF radial variable $\zeta$, where $\zeta^2 = x(1-x)b^2_\perp$ is conjugate to the $q \bar q$ LF kinetic energy $k^2_\perp\over x(1-x)$ for $m_q=0$. This allows the reduction of the dynamics to a single-variable bound state equation acting on the valence $q \bar q$ Fock state.  The confining potential $U(\zeta)$, including its spin-$J$ dependence, is derived from the soft-wall AdS/QCD model with the dilaton  $e^{+\kappa^2 z^2 },$ where $z$ is the fifth coordinate of AdS$_5$ holographically dual  to $\zeta$. See ref.~\cite{Brodsky:2013ar}.   The resulting light-front harmonic oscillator confinement potential $\kappa^4 \zeta^2 $ for light quarks is equivalent to a linear confining potential for heavy quarks in the instant form~\cite{Trawinski:2014msa}. }
\label{FigsJlabProcFig2.pdf}
\end{figure}

A key problem in hadron physics is to obtain a first approximation to QCD which predicts both the hadron spectrum and the hadronic LFWFs.  
If one neglects the Higgs couplings of quarks, then no mass parameter appears in the QCD Lagrangian, and the theory is conformal at the classical level.   Nevertheless,  hadrons have a finite mass.  de Teramond, Dosch, and I~\cite{Brodsky:2013ar}
have shown that a mass gap and a fundamental color confinement scale can be derived from a conformally covariant action when one extends the formalism of de Alfaro, Fubini and Furlan~\cite{deAlfaro:1976je}  to light-front Hamiltonian theory. Remarkably, the resulting light-front potential has a unique form of a harmonic oscillator $\kappa^4 \zeta^2$ in the 
light-front invariant impact variable $\zeta$ where $ \zeta^2Ê = b^2_\perp x(1-x)$. The result is  a single-variable frame-independent relativistic equation of motion for  $q \bar q$ bound states, a ``Light-Front Schr\"odinger Equation"~\cite{deTeramond:2008ht}, analogous to the nonrelativistic radial Schr\"odinger equation in quantum mechanics.  The  Light-Front Schr\"odinger Equation  incorporates color confinement and other essential spectroscopic and dynamical features of hadron physics, including a massless pion for zero quark mass and linear Regge trajectories with the same slope  in the radial quantum number $n$   and internal  orbital angular momentum $L$.   
The same light-front  equation for mesons of arbitrary spin $J$ can be derived~\cite{deTeramond:2013it}
from the holographic mapping of  the ``soft-wall model" modification of AdS$_5$ space with the specific dilaton profile $e^{+\kappa^2 z^2}$,  where one identifies the fifth dimension coordinate $z$ with the light-front coordinate $\zeta$.  The five-dimensional AdS$_5$ space provides a geometrical representation of the conformal group.
It is holographically dual to 3+1  spacetime using light-front time $\tau = t+ z/c$.  
The derivation of the confining LF Schrodinger Equation is outlined in Fig. \ref{FigsJlabProcFig2.pdf}.

The  combination of light-front dynamics, its holographic mapping to AdS$_5$ space, and the dAFF procedure provides new  insight into the physics underlying color confinement, the nonperturbative QCD coupling, and the QCD mass scale.  A comprehensive review is given in ref.~\cite{Brodsky:2014yha}.  The $q \bar q$ mesons and their valence LF wavefunctions are the eigensolutions of a frame-independent bound state equation, the ``Light-Front Schr\"odinger Equation".  The mesonic $q\bar q$ bound-state eigenvalues for massless quarks are $M^2(n, L, S) = 4\kappa^2(n+L +S/2)$.
The equation predicts that the pion eigenstate  $n=L=S=0$ is massless at zero quark mass, The  Regge spectra of the pseudoscalar $S=0$  and vector $S=1$  mesons  are 
predicted correctly, with equal slope in the principal quantum number $n$ and the internal orbital angular momentum.  The predicted nonperturbative pion distribution amplitude 
$\phi_\pi(x) \propto f_\pi \sqrt{x(1-x)}$ is  consistent with the Belle data for the photon-to-pion transition form factor~\cite{Brodsky:2011xx}. The prediction for the LFWF $\psi_\rho(x,k_\perp)$ of the  $\rho$ meson gives excellent 
predictions for the observed features of diffractive $\rho$ electroproduction $\gamma^* p \to \rho  p^\prime$~\cite{Forshaw:2012im}.

\begin{figure}
 \begin{center}
\includegraphics[height=10cm,width=15cm]{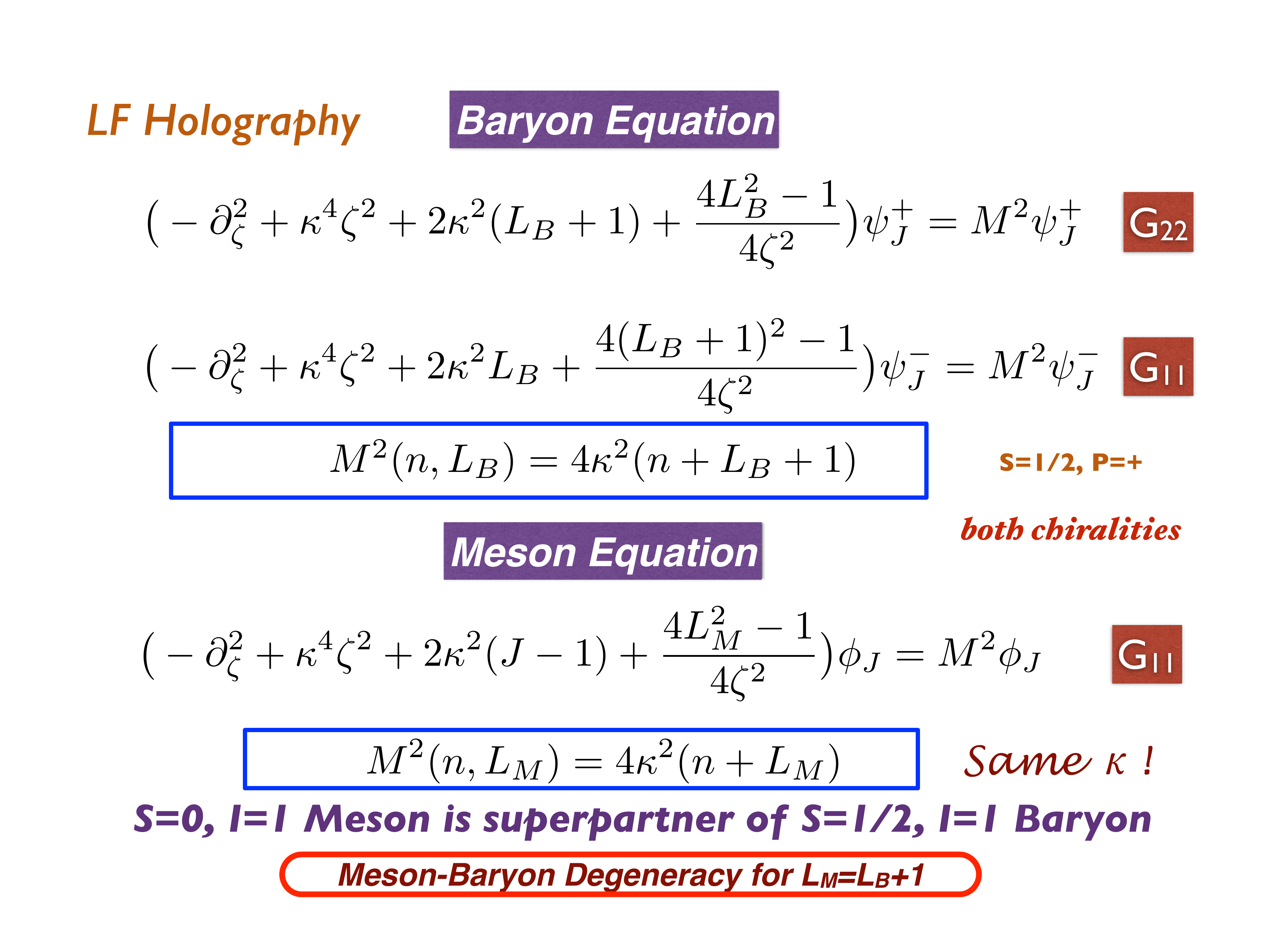}
\includegraphics[height=10cm,width=15cm]{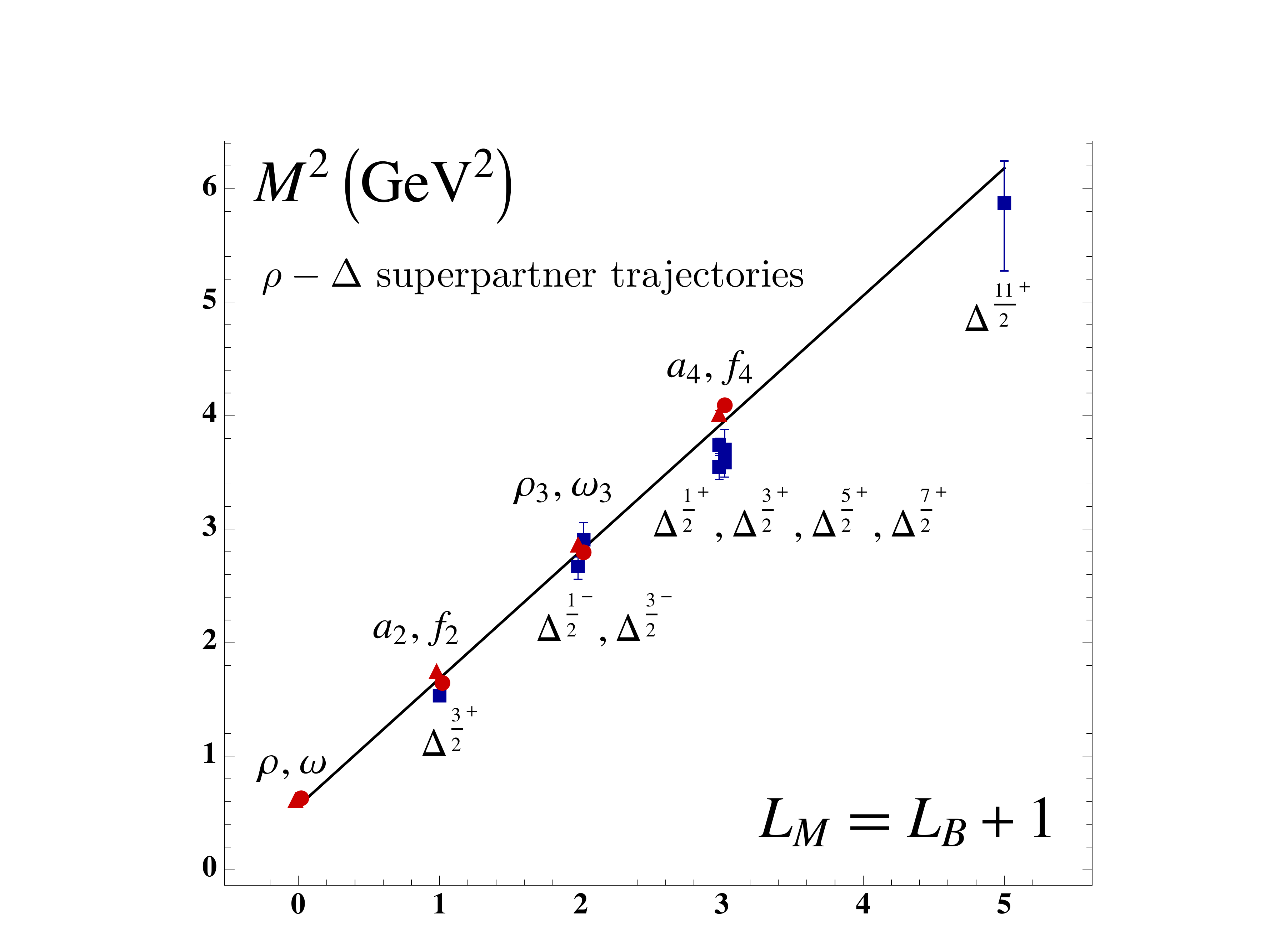}
\end{center}
\caption{(A). The LF Schr\"odinger equations for baryons and mesons for zero quark mass derived from the Pauli $2\times 2$ matrix representation of superconformal algebra.  
The $\psi^\pm$  are the baryon quark-diquark LFWFs where the quark spin $S^z_q=\pm 1/2$ is parallel or antiparallel to the baryon spin $J^z=\pm 1/2$.   The meson and baryon equations are identical if one identifies a meson with internal orbital angular momentum $L_M$ with its superpartner baryon with $L_B = L_M-1.$
See ref.~\cite{deTeramond:2014asa,Dosch:2015nwa,Dosch:2015bca}.
(B). Comparison of the $\rho/\omega$ meson Regge trajectory with the $J=3/2$ $\Delta$  baryon trajectory.   Superconformal algebra  predicts the degeneracy of the  meson and baryon trajectories if one identifies a meson with internal orbital angular momentum $L_M$ with its superpartner baryon with $L_M = L_B+1.$
See refs.~\cite{deTeramond:2014asa,Dosch:2015nwa}.}
\label{FigsJlabProcFig3.pdf}
\end{figure}

These results can be extended~\cite{deTeramond:2014asa,Dosch:2015nwa,Dosch:2015bca} to effective QCD light-front equations for both mesons and baryons by using the generalized supercharges of superconformal algebra~\cite{Fubini:1984hf}. 
The supercharges connect the baryon and meson spectra  and their Regge trajectories to each other in a remarkable manner: each meson has internal  angular momentum one unit higher than its superpartner baryon  $L_M = L_B+1.$  See  Fig. \ref {FigsJlabProcFig3.pdf}(A).   Only one mass parameter $\kappa$ appears; it sets the confinement and the hadron mass scale in the  chiral limit, as well as  the length scale which underlies hadron structure.  ``Light-Front Holography"  not only predicts meson and baryon  spectroscopy  successfully, but also hadron dynamics:  light-front wavefunctions, vector meson electroproduction, distribution amplitudes, form factors, and valence structure functions.  
The LF Schr\"odinger Equations for baryons and mesons derived from superconformal algebra  are shown  in Fig. \ref{FigsJlabProcFig3.pdf}.
The comparison between the meson and baryon masses of the $\rho/\omega$ Regge trajectory with the spin-$3/2$ $\Delta$ trajectory is shown in Fig. \ref{FigsJlabProcFig3.pdf}(B).
Superconformal algebra  predicts the meson and baryon masses are identical if one identifies a meson with internal orbital angular momentum $L_M$ with its superpartner baryon with $L_B = L_M-1.$   Notice that the twist  $\tau = 2+ L_M = 3 + L_B$ of the interpolating operators for the meson and baryon superpartners are the same.   Superconformal algebra also predicts that the LFWFs of the superpartners are identical, and thus they have identical dynamics, such their elastic and transition form factors.   These features can be tested for spacelike  form factors at  JLab12.
\begin{figure}
\begin{center}
\includegraphics[height=10cm,width=15cm]{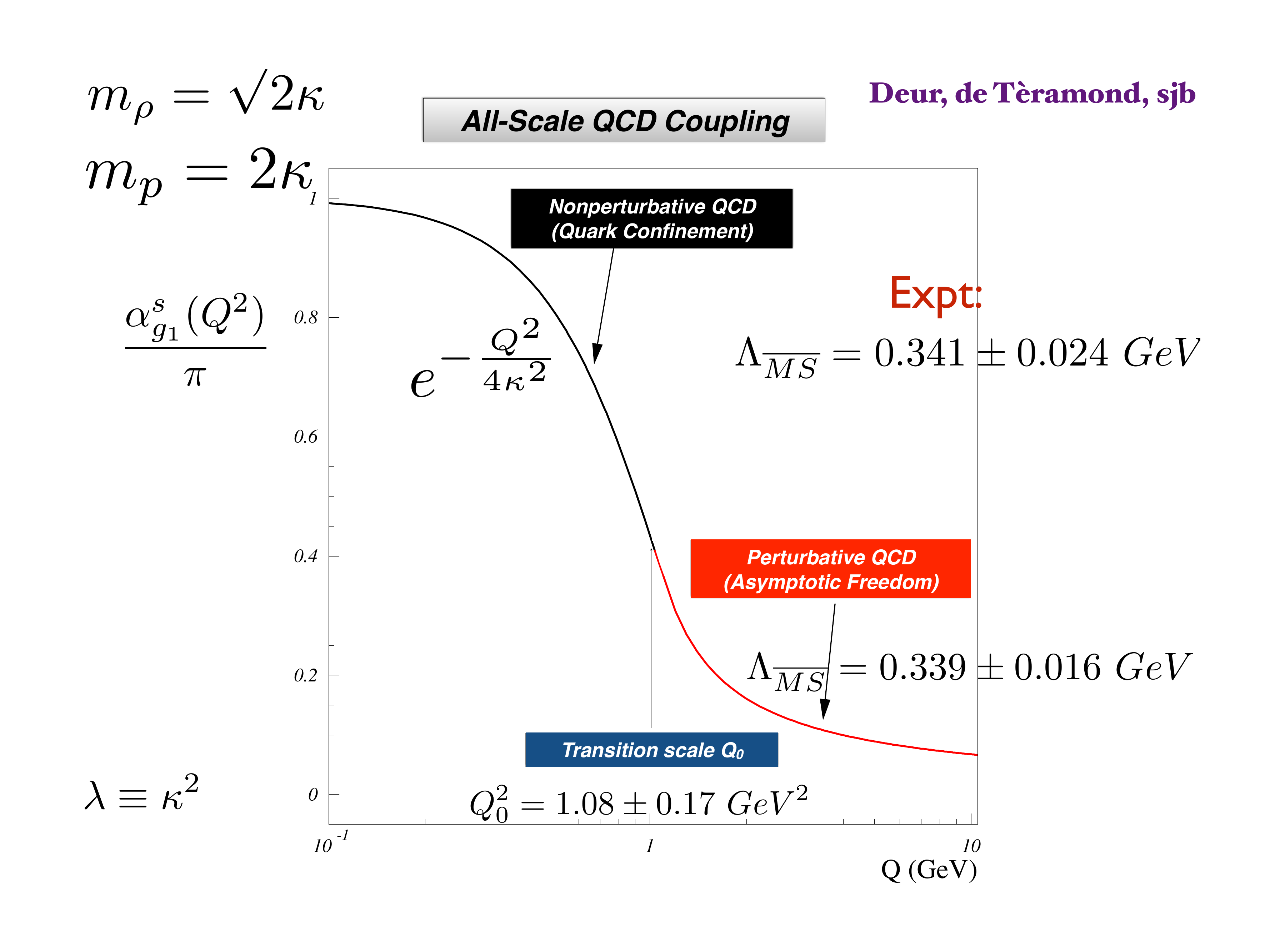}
\includegraphics[height=10cm,width=15cm]{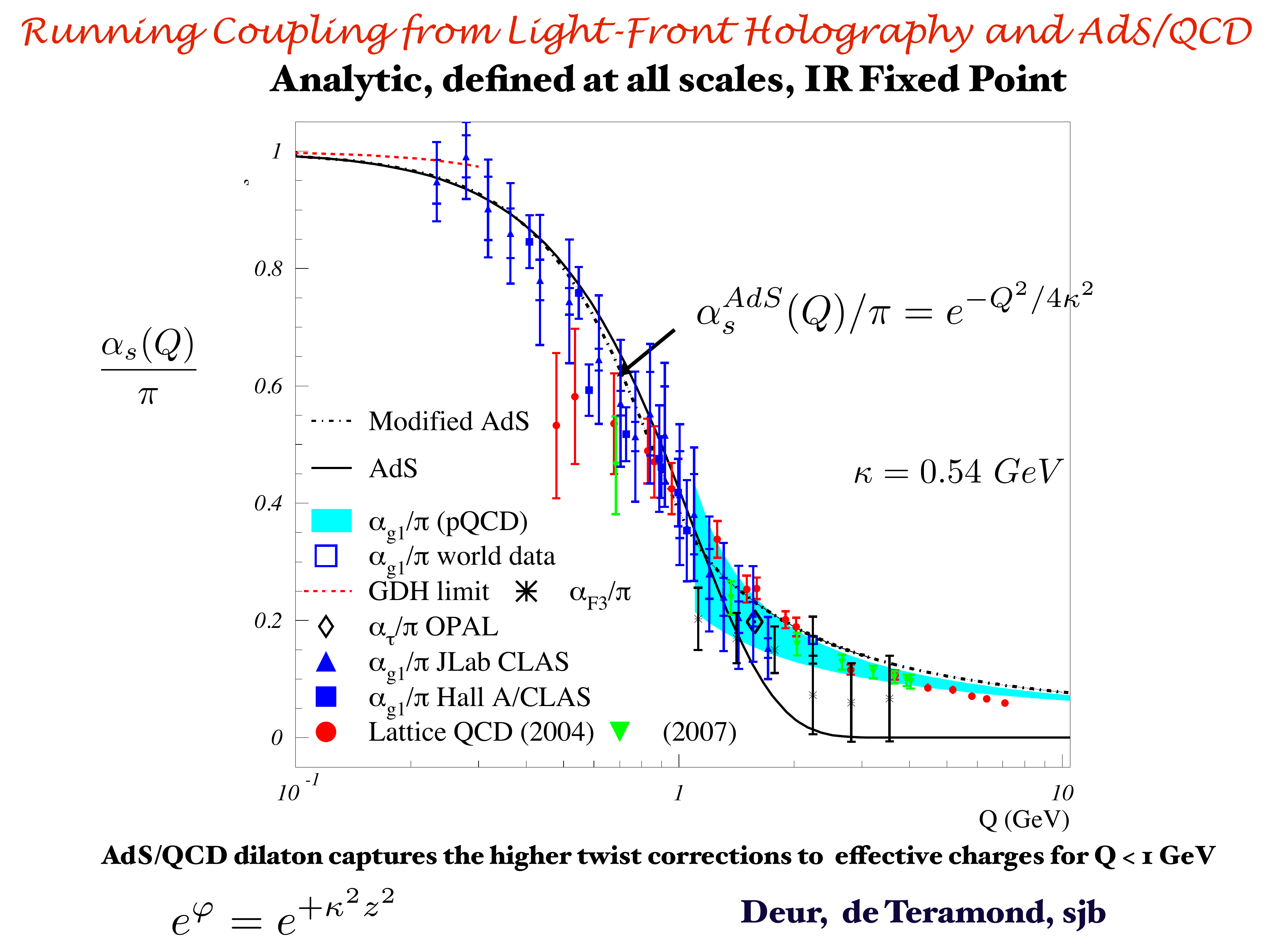}
\end{center}
\caption{
(A)  Prediction from LF Holography for the QCD Running Coupling $\alpha^s_{g_1}(Q^2)$.   The magnitude and derivative of the perturbative and nonperturbative coupling are matched at the scale $Q_0$.  This matching connects the perturbative scale 
$\Lambda_{\overline{MS}}$ to the nonpertubative scale $\kappa$ which underlies the hadron mass scale. 
(B) Comparison of the predicted nonpertubative coupling with measurements of the effective charge $\alpha^s_{g_1}(Q^2)$  
defined from the Bjorken sum rule.  
See ref.~\cite{Brodsky:2014jia}. 
}
\label{FigsJlabProcFig5.pdf}
\end{figure} 
\section {The QCD Coupling at all Scales} 
The QCD running coupling can be defined~\cite{Grunberg:1980ja} at all momentum scales from any perturbatively calculable observable, such as the coupling $\alpha^s_{g_1}(Q^2)$ which is defined from measurements of the Bjorken sum rule.   At high momentum transfer, such ``effective charges"  satisfy asymptotic freedom, obey the usual pQCD renormalization group equations, and can be related to each other without scale ambiguity 
by commensurate scale relations~\cite{Brodsky:1994eh}.  
The dilaton  $e^{+\kappa^2 z^2}$ soft-wall modification of the AdS$_5$ metric, together with LF holography, predicts the functional behavior 
in the small $Q^2$ domain~\cite{Brodsky:2010ur}: 
${\alpha^s_{g_1}(Q^2) = 
\pi   e^{- Q^2 /4 \kappa^2 }}. $ 
Measurements of  $\alpha^s_{g_1}(Q^2)$ are remarkably consistent with this predicted Gaussian form. 
Deur, de Teramond, and I~\cite{Deur:2014qfa,Brodsky:2010ur,Brodsky:2014jia} have also shown how the parameter $\kappa$,  which   determines the mass scale of  hadrons in the chiral limit, can be connected to the  mass scale $\Lambda_s$  controlling the evolution of the perturbative QCD coupling. The connection can be done for any choice of renormalization scheme, such as the $\overline{MS}$ scheme,
as seen in  Fig.~\ref{FigsJlabProcFig5.pdf}. 
The relation between scales is obtained by matching at a scale $Q^2_0$ the nonperturbative behavior of the effective QCD coupling, as determined from light-front holography, to the perturbative QCD coupling with asymptotic freedom.
The result of this perturbative/nonperturbative matching is an effective QCD coupling  defined at all momenta.

\section{Other Features of Light-Front QCD}

There are a number of advantages if one uses  LF Hamiltonian methods for perturbative QCD calculations.  Unlike instant form, where one must sum  $n !$ frame-dependent  amplitudes, only the $\tau$-ordered diagrams where every line has  positive $k^+ =k^0+k^z$  can contribute~\cite{Cruz-Santiago:2015dla}. The number of nonzero amplitudes is also greatly reduced by noting that the total angular momentum projection $J^z = \sum_i^{n-1 } L^z_i + \sum^n_i S^z_i$ and the total $P^+$ are  conserved at each vertex.  In addition, in a renormalizable theory the change in orbital angular momentum is limited to $\Delta L^z =0,\pm 1$ at each vertex.  The calculation of a subgraph of any order in pQCD only needs to be done once;  the result can be stored in a ``history" file, since in LFPth the numerator algebra is independent of the process; the denominator changes, but only by a simple shift of the initial $P^-$. Loop integrations are three dimensional: $\int d^2\vec k_\perp \int^1_0 dx.$
Renormalization can be done using the ``alternate denominator" method which defines the required subtraction counterterms~\cite{Brodsky:1973kb}.

The LF vacuum in LF Hamlitonian theory is defined as the eigenstate of $H_{LF}$ with lowest invariant mass. Since propagation with negative $k^+$  does not appear, there are no loop amplitudes in the LF vacuum -- it is  is thus trivial up to possible $k^+=0$ ``zero"  modes.   The usual quark and gluon QCD vacuum condensates of the instant form =are replaced by physical effects,  such as the running quark mass and the physics contained within the hadronic LFWFs  in the hadronic domain. This is referred to as ``in-hadron" condensates~\cite{Casher:1974xd,Brodsky:2009zd,Brodsky:2010xf}.  In the case of the Higgs theory, the traditional Higgs vacuum expectation value (VEV) is replaced by a zero mode analogous to a classical 
Stark or Zeeman field.~\cite{Srivastava:2002mw}   This again contrasts with the traditional view of the vacuum  based on the instant form. 

The instant-form vacuum, the lowest energy eigenstate of the instant-form Hamiltonian,  is defined at one instant of time over all space; it is thus acausal and frame-dependent.  It is usually argued that the QCD contribution to the cosmological constant -- dark energy  -- is $10^{45}$ times larger that observed, and in the case of the Higgs theory, the Higgs VEV is argued to be $10^{54}$ larger than observed~\cite{Zee:2008zz}, 
estimates based on the loop diagrams of the acausal frame-dependent instant-form vacuum.  However, the universe is observed within the causal horizon, not at a single instant of time.  In contrast, the light-front vacuum provides a viable description of the visible universe~\cite{Brodsky:2010xf}. Thus in agreement with Einstein, quantum effects do not contribute to the cosmological constant.   In the case of the HIggs theory, the  Higgs zero mode has no energy density,  so again  it gives no contribution to the cosmological constant.  However, it is possible that if one solves the Higgs theory in a curved universe, the zero mode will be replaced with a field of nonzero curvature which could give a nonzero contribution.

\section{Is the Momentum Sum Rule Valid for Nuclear Structure Functions? }

Sum rules for DIS  processes are analyzed using the operator product expansion of the forward virtual Compton amplitude, assuming it depends in the limit $Q^2 \to \infty$ on matrix elements of local operators such as the energy-momentum tensor.  The moments of the structure function and other distributions can then be evaluated as overlaps of the target hadron's light-front wavefunction,  as in the Drell-Yan-West formulae for hadronic form factors~\cite{Brodsky:1980zm,Liuti:2013cna,Mondal:2015uha,Lorce:2011dv}.
The real phase of the resulting DIS amplitude and its OPE matrix elements reflects the real phase of the stable target hadron's wavefunction.

The ``handbag" approximation to deeply virtual Compton scattering also defines the ``static"  contribution~\cite{Brodsky:2008xe,Brodsky:2009dv} to the measured parton distribution functions (PDF), transverse momentum distributions, etc.  The resulting momentum, spin and other sum rules reflect the properties of the hadron's light-front wavefunction.
However, final-state interactions which occur {\it after}  the lepton scatters on the quark, can give non-trivial contributions to deep inelastic scattering processes at leading twist and thus survive at high $Q^2$ and high $W^2 = (q+p)^2.$  For example, the pseudo-$T$-odd Sivers effect~\cite{Brodsky:2002cx} is directly sensitive to the rescattering of the struck quark. 
Similarly, diffractive deep inelastic scattering involves the exchange of a gluon after the quark has been struck by the lepton~\cite{Brodsky:2002ue}.  In each case the corresponding DVCS amplitude is not given by the handbag diagram since interactions between the two currents are essential.
These ``lensing" corrections survive when both $W^2$ and $Q^2$ are large since the vector gluon couplings grow with energy.  Part of the phase can be associated with a Wilson line as an augmented LFWF~\cite{Brodsky:2010vs} which do not affect the moments.  

The Glauber propagation  of the vector system $V$ produced by the diffractive DIS interaction on the nuclear front face and its subsequent  inelastic interaction with the nucleons in the nuclear interior $V + N_b \to X$ occurs after the lepton interacts with the struck quark.  
Because of the rescattering dynamics, the DDIS amplitude acquires a complex phase from Pomeron and Regge exchange;  thus final-state  rescattering corrections lead to  nontrivial ``dynamical" contributions to the measured PDFs; i.e., they involve physics aspects of the scattering process itself~\cite{Brodsky:2013oya}.  The $ I = 1$ Reggeon contribution to diffractive DIS on the front-face nucleon leads to flavor-dependent antishadowing~\cite{Brodsky:1989qz,Brodsky:2004qa}.  This could explain why the NuTeV charged current measurement $\mu A \to \nu X$ scattering does not appear to show antishadowing
 in contrast to deep inelastic electron nucleus scattering as discussed in ref. ~\cite{Schienbein:2007fs}.
Again the corresponding DVCS amplitude is not given by the handbag diagram since interactions between the two currents are essential.

Diffractive DIS is leading-twist and is the essential component of the two-step amplitude which causes shadowing and antishadowing of the nuclear PDF.  It is important to analyze whether the momentum and other sum rules derived from the OPE expansion in terms of local operators remain valid when these dynamical rescattering corrections to the nuclear PDF are included.   The OPE is derived assuming that the LF time separation between the virtual photons in the forward virtual Compton amplitude 
$\gamma^* A \to \gamma^* A$  scales as $1/Q^2$.
However, the propagation  of the vector system $V$ produced by the diffractive DIS interaction on the front face and its inelastic interaction with the nucleons in the nuclear interior $V + N_b \to X$ are characterized by a longer LF time  which scales as $ {1/W^2}$.  Thus the leading-twist multi-nucleon processes that produce shadowing and antishadowing in a nucleus are evidently not present in the $Q^2 \to \infty$ OPE analysis.

It should be emphasized  that shadowing in deep inelastic lepton scattering on a nucleus  involves  nucleons at or near the front surface; i.e, the nucleons facing the incoming lepton beam. This  geometrical orientation is not built into the frame-independent nuclear LFWFs used to evaluate the matrix elements of local currents.  Thus the dynamical phenomena of leading-twist shadowing and antishadowing appear to invalidate the sum rules for nuclear PDFs.  The same complications occur in the leading-twist analysis of deeply virtual Compton scattering $\gamma^* A \to \gamma^* A$ on a nuclear target.

\section{Elimination of  Renormalization Scale Ambiguities}
The ``Principle of Maximum Conformality", (PMC)~\cite{Wu:2013ei} systematically eliminates the renormalization scale ambiguity in perturbative QCD calculations, order-by-order.    The PMC predictions are also insensitive to the choice of the initial renormalization scale $\mu_0.$
The PMC sums all of the non-conformal terms associated with the QCD $\beta$ function into the scales of the coupling at each order in pQCD.
The resulting  conformal series is free of renormalon resummation problems.  The number
of active flavors $n_f$ in the QCD $\beta$ function is also
correctly determined at each order.  
The $R_\delta$ scheme -- a generalization  of t'Hooft's  dimensional regularization. systematically  identifies the nonconformal $\beta$ contributions to any perturbative QCD series, thus allowing the automatic implementation of the PMC procedure~\cite{Mojaza:2012mf}.     
 The resulting scale-fixed predictions for physical observables using
the PMC are {\it  independent of
the choice of renormalization scheme} --  a key requirement of 
renormalization group invariance.  The PMC provides a generalization of  the BLM method~\cite{Brodsky:1982gc} to all orders in pQCD.  
A related approach is given in refs.~\cite{Kataev:2014zha,Kataev:2014jba,Kataev:2014zwa}.
 
The elimination of the renormalization scale ambiguity greatly increases the precision, convergence, and reliability of pQCD predictions.  
For example, PMC scale-setting has been applied to the pQCD prediction for $t \bar t$ pair production at the LHC,  where subtle aspects of the renormalization scale of the three-gluon vertex and multi-gluon amplitudes, as well as  large radiative corrections to heavy quarks at threshold play a crucial role.  
The large discrepancy of pQCD predictions with  the $t \bar t$  forward-backward asymmetry measured at the Tevatron is significantly reduced from 
$3 \sigma$ to approximately $ 1 \sigma$~\cite{Brodsky:2012rj,Brodsky:2012sz}.

\section*{Acknowledgements}

Presented at the 18th Workshop, ``What Comes Beyond the Standard Models", Bled, Slovenia July 11-19, 2015. 
I thank my collaborators, James Bjorken, Kelly Chiu, Alexandre Deur, Guy de Teramond, Guenter Dosch, Susan Gardner, Fred Goldhaber,  Paul Hoyer, Dae Sung Hwang,  Rich Lebed, 
Simonetta Liuti, Cedric Lorce, Matin Mojaza,  Michael Peskin, Craig Roberts, Robert Schrock, Ivan Schmidt, Peter Tandy, and Xing-Gang Wu.
 for helpful conversations and suggestions.
This research was supported by the Department of Energy,  contract DE--AC02--76SF00515.  
SLAC-PUB-16432.

\end{document}